\magnification = \magstep1
\baselineskip = 15pt
\parskip = 4pt

\newcount\ftnumber
\def\ft#1{\global\advance\ftnumber by 1
          {\baselineskip=12pt \footnote{$^{\the\ftnumber}$}{#1 }}}
\def\section#1{\vskip 12pt \centerline{{\sl #1}} \vskip 6pt}
\font\title = cmr10 scaled 1440

\centerline{{\title Mysl, smysl, sv\u et}}

\vskip 15pt

\centerline{N. David Mermin}
\vskip10pt

\centerline{Acceptance speech}

\centerline{Dagmar and V\'aclav Havel Foundation VIZE 97 Prize for 2017}

\centerline{Prague, October 5, 2017}

\vskip 15pt

Twelve years ago, on the hundredth anniversary of Einstein's first paper on relativity, I published a book for the general reader.\ft{{\it It's about Time: Understanding Einstein's Relativity\/}, Princeton UP, 2005.}   At the very end of the book I say this:\vskip 3pt
{\narrower \narrower

\noindent  \sl What makes the pursuit of science so engrossing is to learn that one's most strongly held beliefs can be completely wrong.  The search to identify and correct the old errors can lead to deep insights into nature. {\sl The world would be a far better place for all of us if this joy scientists find in exposing their own misconceptions were more common in other areas of human endeavor.}

}
\vskip 5pt
  
When I wrote this in 2005 I was thinking of  President George W. Bush.   In preparing this speech I discovered that your 2005 Laureate, Philip G. Zimbardo, had similar ideas about Bush.  Zimbardo said:

{\narrower \narrower
 
\noindent \sl  [The Czech] presidential flag bears the motto, {\it pravda v\'{\i}te\u zi\/},  ``Truth Prevails"!   I am sorry to say that such a banner is unlikely to fly high over the offices of the current administration in the United States.

}

\noindent Neither of us could have imagined then what would be happening in America 12 years later.

\vskip 10pt 

In 2006, Vacl\'av Havel himself articulated the nature of America's current problem.  Asked about lessons he had learned from his presidency, Havel said:\ft{{\it Prosim Stru\u cn\u e\/} (please be brief).  Page references are to the English translation: {\it To the Castle and Back\/}, Vintage Books, 2008.}

{\narrower \narrower

\noindent  \sl People expect their head of state to declare the importance of certain moral norms. [p. 171]  

} 
\noindent And he said: 

{\narrower \narrower

\noindent  \sl Again and again I realized how important it is to have a very ordinary thing: good taste.  [p. 323]  

}

\vskip 15pt
\hskip -2pt And early in the Trump administration, your 2015 Laureate, Timothy Snyder, thinking of the {\it Reichstag\/} fire, issued a warning:\ft{Interview in {\it Suddeutsche Zeitung\/}, February 7, 2017.}

{\narrower \narrower
\noindent   \sl If a terror attack happens in the United States $\ldots$ it is not a reason to suspend  rights or declare a state of emergency.   History teaches us the tricks of authoritarians.  We cannot allow ourselves to fall for them.

}

\vskip 10pt  
So today I'm proud to accept an award established by President and Mrs.~Havel.  And I'm proud to join a group that includes Professors Zimbardo and  Snyder.  
\vskip 12pt
\centerline{***}
\vskip 12pt
But I'm a physicist and I want to talk today about wrong beliefs in physics. 
Fifty years after his great 1905 paper, Einstein put the wrong belief exposed by relativity this way:\ft{J. R. Shankland, ``Conversations with Albert Einstein", American Journal of Physics, {\bf 31}, 47-57 (1963).}  

{ \narrower \narrower

\noindent  \sl At last it came to me that time was suspect!   

}
\noindent What he meant was that certain troubles in electromagnetism all stemmed from a misunderstanding of  time itself. 
Today we know that clocks are not devices we have invented to help us measure a  thing called ÒtimeÓ.    Time is a concept we have invented to help us make sense of relations that exist between the things we call clocks.   

An extreme version of this new understanding of time can be found in my favorite of all the wonderful quotations attributed to Einstein himself:

{\narrower \narrower

\noindent  \sl Space and time are modes in which we think, not conditions in which we live.

}

This is as lovely a piece of poetry as Einstein ever produced. But I doubt he ever
said it. It seems to come from a biography that cites it only as a private remark.\ft{Remark to Paul Ehrenfest,  cited without a source by Aylesa Forsee in {\it Albert Einstein: Theoretical Physicist\/}, MacMillan, N.Y. (1963), p.~81.}  If Einstein  had
believed that space and time were modes of thought and not features of the world, I doubt
he would have found quantum mechanics as unacceptable as he famously did.
 
Which brings me to my primary topic.

\vskip 12pt
\centerline{***}
\vskip 12pt

Quantum mechanics is the theory of matter developed during the first quarter of the 20th century.   
It has transformed the nature of science, and the discoveries it has lead to have changed the way we live.
But it  is very,  very peculiar.  

For over 90 years there has been no agreement  about the meaning of the concepts used by the quantum theory.    Unlike most philosophical issues raised by physics, the question raised by quantum mechanics is not a hair-splitting refinement.  The question is ``What the hell are we talking about?''   

I started my career as a quantum engineer.    I am coauthor of a well-known book on solid-state physics.  Solid state physics is  applied quantum mechanics.   However,  my ultimate aim as a physicist has always been to understand just what it was that I was actually doing.  

Five years ago, I realized that two younger colleagues, Chris Fuchs and R\"udiger Schack, had a new way of thinking about science and quantum mechanics that actually made sense.   ``What are {\it we\/} talking about?'' is the wrong question.    The right question is ``What am {\it I\/} talking about when {\it I \/} use quantum mechanics?''  

Any user of quantum mechanics can ask this question.   The answer, for each of us, is that I am talking about my personal experience.  When I take an action on the world, quantum mechanics tells me  the likelihood of the kind of experience the world will induce back in me, in response to my action.  ``Likelihood'' means my personal expectation, based on how the world has responded to me in the past.

Quantum mechanics is a  personal guide to betting.   Placing bets famously bothered Einstein, but today's physicists have no trouble with God playing dice.  What does trouble them is the idea that  quantum mechanics might be a {\it personal\/} guide, producing different expectations, when used by different people with different experiences.   Nine decades of controversy and confusion have emerged from unsuccessful efforts to avoid the personal by reifying individual human expectations and experience into objective facts about the inanimate world. 
 
This smells of solipsism.    My experience is private to me.    Yours is private to  you.   Are each of us therefore locked into our own unavoidably private worlds?    No, we are not, because of language.  

Language enables  each of us to give  verbal representations of our own private experience to other people.  My experience of your words is not identical to the experience you are trying to represent, but I have learned to infer from your words that you are able to feel something analogous to what  I experience.   Often your verbal representations are crude, but sometimes, in your books or scientific articles, they can be quite refined.   Language enables us collectively to come to something like a  common understanding of the character of the worlds that  each of us constructs from our own private experience.  

Niels Bohr--- the dominant figure of the first quantum generation ---  always stressed the importance of being able to state the outcomes of experiments in ``ordinary language''.  I never understood why.    Now, thanks to Fuchs and Schack, I do.   Language is the closest you and I can get to comparing your private experience with mine.  

I don't think that's what Bohr had in mind.   When he used the term ``our experience'' he used the first person plural to mean ``all of us'' --- not ``each one of us''.    He did not view language as an attempt to overcome the inherent privacy of personal experience.   By ``ordinary language'' I think he meant a description of the instruments we invent to help probe a world not directly accessible to our senses.  I believe Bohr, like all physicists, was trying to make comfortably external and objective what is uncomfortably  internal and subjective. 

This dehumanization of both the act and the content of observation is taken for granted by physicists.  Our refusal to acknowledge the subjective, personal side of science, lies behind ninety years of controversy about quantum mechanics.    Steven Weinberg --- one of the leading physicists of my own generation --- made the prevailing view explicit earlier this year, when he wrote in the {\it New York Review\/} that he\ft{New York Review of Books, January 19, 2017. See also Letters, April 6, 2017.} 

{\narrower \narrower

\noindent  \sl hoped for a physical theory that would allow us to deduce what happens  $\ldots$ from impersonal laws that apply to everything, without giving any special status to people in these laws.

}


That ``special status to people'' lies only in the particular circumstances under which the impersonal laws are applied.    I use the same laws of quantum mechanics as anybody else.  I use them to infer from the contents of my past experience, the likelihood of my subsequent experience, as a guide to my subsequent action.  The impersonal laws are designed for anybody who has learned enough physics to use them.

Everyone would agree that people play an essential role in the construction of physical science, but most physicists maintain that people can then be removed from the story, as  scaffolding is taken away from a finished building.   The lesson of quantum mechanics is that if you want to make sense of what you are talking about, then you cannot  ignore the fact that it is people who use those laws of physics.

\vskip 15pt

In popular scientific writings  --- what Einstein called\ft{``Einstein Attacks Quantum Theory'' {\it New York Times\/}, May 4, 1935.}   the ``secular press'' --- a common misreading of Fuchs and Schack has been ``It's all mind.''   This is as wrong as the prevailing opinion among physicists that it's all world. There is 
mind and there is world. Quantum mechanics has taught us that we cannot understand 
what we are talking about without invoking both. What links the contents of the mind to the world 
that induces those contents  is the meaning each of us finds in our own experience of that world.

Each Vize 97 laureate prepares a book of essays in Czech.  Karl Pribram began the series  in 1999 with  {\it Mozek a mysl\/} --- brain and mind.  Umberto Eco followed him with  {\it Mysl a smysl\/} --- mind and meaning.  Zden\u ek Neubauer then completed the series with {\it  Smysl a sv\u et\/} --- meaning and world.  After those first three the pattern was broken.   But Pribram, Eco, and Neubauer have given me a most appropriate title:  {\it Mysl, smysl, sv\u et.\/}  

If I had to make a coat of arms for Fuchs' and Schack's insight into quantum mechanics,  it would carry the words: {\it mysl, smysl, sv\u et.\/}   They would have to be in Czech. ``Mind, meaning, world'' has no poetry in it.    And what  physicists' understanding of quantum mechanics has lacked, for ninety years, is poetry.

\bye